\documentclass[trackchanges]{aastex701}

\newcommand{\tamojeet}[1]{{\color{black} #1}}
\newcommand{\Sg}{Sgr~A*}
\newcommand{\Ms}{Mini-spiral}
\newcommand{\Msol}{\hbox{M$_\odot$}}
\newcommand{\DEL}[1]{}
\newcommand{\0}{\phantom{0}}

\usepackage{amsmath}
\usepackage{graphicx}
\usepackage{braket}
\usepackage{xcolor}
\usepackage{txfonts}

\begin{document}

   \title{Photometric Constraints on Intermediate-mass Black Holes 
   in the Galactic Centre}

    \author[0009-0003-9906-2745]{Tamojeet Roychowdhury}
    \affiliation{Department of Astronomy, University of California Berkeley, Berkeley, CA 94704, USA}
    \email[show]{tamojeet@berkeley.edu}

    \author[orcid=0000-0002-9156-2249,sname='von Fellenberg', gname=Sebastiano]{Sebastiano D. von Fellenberg}
    \altaffiliation{Feodor Lynen Fellow}
    \affiliation{Canadian Institute for Theoretical Astrophysics, University of Toronto, 60 St.\ George Street, Toronto, ON M5S 3H8, Canada}
    \affiliation{Max Planck Institute for Radioastronomy, auf dem H{\"u}gel 69, Bonn, Germany }
    \email{sfellenberg@cita.utoronto.ca}

    \author[orcid=0000-0003-3503-3446,sname=Michail, gname=Joseph]{Joseph M. Michail}
    \altaffiliation{NSF Astronomy \& Astrophysics Postdoctoral Fellow}
    \affiliation{Center for Astrophysics $|$ Harvard \& Smithsonian, 60 Garden Street, Cambridge, MA 02138, USA}
    \email{joseph.michail@cfa.harvard.edu}  

    \author[0000-0002-9895-5758]{S. P. Willner}
    \affiliation{Center for Astrophysics $|$ Harvard \& Smithsonian, 60 Garden Street, Cambridge, MA 02138, USA}
    \email{swillner@cfa.harvard.edu}

    \author[0000-0001-8921-3624]{Nicole M. Ford}
    \affiliation{McGill University, Montreal QC H3A 0G4, Canada}%
    \affiliation{Trottier Space Institute, 3550 Rue University, Montréal, Québec, H3A 2A7, Canada}
    \email{nicole.ford@mail.mcgill.ca}

    \author[0009-0004-8539-3516]{Zach Sumners}
    \affiliation{McGill University, Montreal QC H3A 0G4, Canada}%
    \affiliation{Trottier Space Institute, 3550 Rue University, Montréal, Québec, H3A 2A7, Canada}
    \email{nicole.ford@mail.mcgill.ca}

    \author[]{Sophia Sanchez-Maes}
    \affiliation{University of Maryland, College Park, MD 20742, USA}
    \email{sophiasm@umd.edu}
    
    \author[0000-0001-9554-6062]{Tuan Do}
    \affiliation{Department of Physics \& Astronomy,  University of California, Los Angeles, 90095-1547, USA}
    \email{tdo@astro.ucla.edu}

    \author[0000-0003-4801-0489]{Macarena Garcia Marin}
    \affiliation{European Space Agency (ESA), ESA Office, Space Telescope Science Institute, 3700 San Martin Drive, Baltimore, MD 21218, USA}
    \email{Macarena.Garcia.Marin@esa.int}

    \author[0000-0001-9564-0876]{Sera Markoff}   \affiliation{Anton Pannekoek Institute for Astronomy, University of Amsterdam, Science Park 904, 1098 XH Amsterdam, The Netherlands}   \affiliation{Gravitation and Astroparticle Physics Amsterdam Institute, University of Amsterdam, Science Park 904, 1098 XH 195 196 Amsterdam, The Netherlands}
  \email{S.B.Markoff@uva.nl}
    
    \author[0000-0002-0670-0708]{Giovanni G. Fazio}
    \affiliation{Center for Astrophysics $|$ Harvard \& Smithsonian, 60 Garden Street, Cambridge, MA 02138, USA}
    \email{gfazio@cfa.harvard.edu}
    
    \author[0000-0001-6803-2138]{Daryl Haggard}
    \affiliation{McGill University, Montreal QC H3A 0G4, Canada}%
    \affiliation{Trottier Space Institute, 3550 Rue University, Montréal, Québec, H3A 2A7, Canada}
    \email{daryl.haggard@mcgill.ca}
    
    \author[0000-0002-5599-4650]{Joseph L. Hora}
    \affiliation{Center for Astrophysics $|$ Harvard \& Smithsonian, 60 Garden Street, Cambridge, MA 02138, USA}
    \email{jhora@cfa.harvard.edu}
    
    \author[0000-0002-7301-3908]{Bart Ripperda}
    \affiliation{Canadian Institute for Theoretical Astrophysics, University of Toronto, 60 St.\ George Street, Toronto, ON M5S 3H8, Canada}
    \affiliation{Dunlap Institute for Astronomy and Astrophysics, University of Toronto, 50 St.\ George Street, Toronto, ON M5S 3H4, Canada}
    \affiliation{Department of Physics, University of Toronto, 60 St. George Street, Toronto, ON M5S 1A7, Canada.}
    \email{bartripperda@gmail.com}

    \author[]{Nadeen B. Sabha}
    \affiliation{University of Innsbruck, Institut für Astro- und Teilchenphysik, Technikerstr. 25/8, 6020 Innsbruck, Austria}
    \email{Nadeen.Sabha@uibk.ac.at}
    
    \author[]{Howard A. Smith}
    \affiliation{Center for Astrophysics $|$ Harvard \& Smithsonian, 60 Garden Street, Cambridge, MA 02138, USA}
    \email{hsmith@cfa.harvard.edu}
    
    \author[0000-0003-2618-797X]{Gunther Witzel}
    \affiliation{Max Planck Institute for Radioastronomy, auf dem H{\"u}gel 69, Bonn, Germany }
    \email{gwitzel@mpifr-bonn.mpg.de}

\date{Received ; accepted }

   \begin{abstract}
   JWST/MIRI observations can place photometric limits on the presence of an intermediate-mass black hole (IMBH) near the Galactic Centre. The stellar complex IRS~13E,  a co-moving conglomerate of young and massive stars, is a prime location to study because it has been speculated to be bound by an IMBH\null. Assuming a standard radiatively inefficient accretion flow (RIAF) and a minimum fractional variability of 10\% of intrinsic luminosity, the wavelength of peak emission in the spectral energy distribution for an IMBH would lie in the mid-infrared  ($\sim$5--25~$\mu$m),  and variability would be detectable in MIRI time-series observations. Monitoring fails to detect such variable emission (other than from \Sg) in and around the IRS~13E complex, and upper limits on a putative IMBH's intrinsic variability on timescales of minutes to $\sim$1~hour are $\lesssim$1~mJy at 12~$\mu$m and $\lesssim$2~mJy at 19~$\mu$m. These translate to luminosities ${\lesssim}25\times 10^{32}$~erg~s$^{-1}$. The resulting limits on the  IMBH mass and accretion rate rule out any IMBH with mass ${\gtrsim} 10^3$~\Msol\ accreting at ${\gtrsim} 10^{-6}$ times Eddington rate at the location of IRS~13E\null.Further, the observations rule out an IMBH anywhere in the central $6''\times 6''$ region that is more massive than ${\approx} 2\times10^{3}$~\Msol\ and accreting at ${\geq} 10^{-6}$ of the Eddington rate.
   Assuming Bondi accretion scaled to typical RIAF-accretion efficiencies, albeit somewhat uncertain, also allows us to rule out IMBHs moving with typical velocities ${\sim}200~\mathrm{km~s^{-1}}$ and masses ${\gtrsim} 2\times10^3$~\Msol\. These methods showcase the effectiveness of photometric variability measurements in constraining the presence of accreting black holes in Galactic centre-like environments.

\end{abstract}

  
\section{Introduction}
The Galactic Centre (GC), located 8178 parsecs away \citep{GRAVITYCollaboration2019}, harbours a massive black hole (MBH) with a mass of ${\approx} 4\times10^6~\mathrm{M_{\odot}}$ \citep[e.g.,][]{GravityCollaboration2021_abberations,Do2019}. 
Robust observations of the motion of stars in the black hole's close proximity \citep[e.g.,][]{Ghez1998, Schodel2002} and the black hole's association with an X-ray \citep[e.g.,][]{Baganoff2001_flare, Eckart2006, Ponti2017,GravityCollaboration2021_xrayflare}, an infrared \citep[e.g.,][]{Genzel2003,Do2009,vonFellenberg2023_sgra}, and a radio source \citep[\Sg, e.g.,][]{Balick1974,Bower2003,Falcke2000_bhshadow,eht_sgra_I} have established 
several key observational aspects of \Sg\ and its surrounding star cluster.

The GC Nuclear Star Cluster is isotropic, rotates in the same orientation as the Galaxy, extends to $\sim$60--100\arcsec\ ($\sim$2.4--4 parsecs), and has approximately the same mass as the MBH (6--$9\times 10^6$~\Msol, e.g., \citealt{Chatzopoulos2015, Fritz2016, FeldmeierKrause2025}). The observable stars are predominantly metal-rich, old, and giant-type \citep{Schodel2010,Schoedel2018}.  Most of them formed with the Galaxy, but $\sim$20\% formed in several later, distinct star-formation events \citep{Pfuhl2011, Schoedel2020},
the last of which occurred approximately 4--6~{Myr} ago \citep[]{Martins2007, Do2013}. These recently-formed stars, generally referred to as ``the young stars,'' show a distinct dynamical configuration in at least two stellar disks \citep{Paumard2006, Lu2006, Lu2009, Bartko2009, Yelda2014,vonFellenberg2022,Jia2023}. The young stars most likely formed in a massive accretion disk when its vertical collapse induced star formation \citep[e.g.,][]{Bonnell2008,HobbsNayakshin2009}. Alternative scenarios, such as the in-fall of a young star cluster, are increasingly disfavored due to the absence of a trail of young stars outside the Nuclear Star Cluster \citep{Feldmeier-Krause2015,FeldmeierKrause2025_spec}. Similarly, the rejuvenation of older stars, either through stellar collision or accretion, is challenged by the seemingly normal spectra of the young stars \citep{Habibi2017}. In the inner-most region ($r<1''$), the dynamical properties of the young stars change from a predominantly coherent, disk-like structure to a predominantly isotropic distribution \citep{Ghez2008,Gillessen2009,Boehle2016,Gillessen2019} but possibly with two edge-on disks resultant of a disturbance at an IRS 13E--like distance \citep{Ali_2020}.

The possible presence of an intermediate-mass black hole (IMBH) in the GC has received considerable attention.
An IMBH was initially proposed as a potential solution to account for the presence of young stars in the GC \citep[the in-spiral scenario:][]{PortegiesZwart2003}. IRS~13E was proposed as a potential remnant cluster \citep[e.g.,][]{Maillard2004} composed of co-moving young stars held together by the presence of an ${\sim} 10^4~\mathrm{M_\odot}$ IMBH \citep[e.g.,][]{Schoedel2005}.
While the in-spiral scenario is now considered observationally unlikely, as mentioned above, the  presence of an IMBH in IRS~13E remains a possibility. 
Indirect observational evidence in favor of an IMBH comes from accelerated gas motions in its vicinity \citep[e.g.,][]{Tsuboi2017}, but those may also be explained by colliding winds of the Wolf--Rayet (WR) stars present \citep{Wang2020,Zhu2020}. The X-ray detection of the cluster may also be caused by the interaction of WR winds \citep[e.g.,][]{Baganoff2003_spectrum} rather than accretion onto a putative IMBH\null.
In addition, a statistical analysis of the stellar distribution in the GC shows that a chance occurrence of an IRS~13E--like association (i.e., three massive early-type stars observed in the vicinity of each other) is plausible \citep{Fritz2010}, especially as the IRS~13E stars may be associated with one of the young stellar disks \citep{Jia2023}. 

Despite the absence of conclusive evidence, the presence of a GC IMBH in IRS~13E remains a widely discussed possibility, especially as a convenient solution for the observed kinematic properties of young stars within $1''$ of \Sg, loosely referred to as the ``S-stars'' \citep[e.g.,][]{Ali_2020, Gillessen2009}.
Assuming these stars originated from the same stellar disk(s) as the young stars at larger radii, their dynamical configuration is difficult to explain. While dynamical processes can relax a coherent angular-momentum distribution, no known process can do so on timescales as short as $\sim$6~{Myr} (e.g., \citealt{Kocsis2011}, but \citealt{Chen2014} suggested a possible mechanism). Three-body interactions between a stellar binary and the MBH can efficiently create a random angular momentum distribution  through the Hills mechanism \citep{Hills1988}, but the observed eccentricities are too low to be solely attributable to that. (\citealt{Gillessen2017} and \citealt{Alexander2017} provide reviews.)
The presence of the IMBH would significantly alter the dynamical properties of the system, as the stars are bound to the central MBH but also perturbed by the IMBH, allowing  more flexibility to explain the observed stellar distribution \citep[e.g.,][]{Zheng2020,Zheng2021,Burkert2024}.

Different avenues exist to rule out an IMBH in the GC\null. Most methods rely on the motion of the stars to constrain the allowed parameter space \citep{Hansen2003,Yu2003,Gillessen2009,Naoz2020,GravityCollobration2023_IMBH, Will2023} or the reflex motion of \Sg\ in radio very long baseline interferometry (VLBI) observations \citep{Reid2004, Reid2020}. 

This paper explores a novel avenue to constrain the presence of the IMBH: direct photometric constraints based on the absence of significant variability in the {\em infrared} light curve of IRS~13E and other regions of the central few parsecs. (\citealt{Fritz2010} suggested the absence of X-ray variability made an IMBH unlikely but gave no quantitative limits.) We assume that the emission of a putative IMBH is similar to that of \Sg---a so-called radiative inefficient accretion flow \citep[RIAF, e.g.,][]{Narayan1994,Abramowicz1995,Blandford1999,Quataert2000}---and therefore we can use the observed properties and models of \Sg\ to guide the expected properties of the IMBH.

This paper is organized as follows: Sec.~2 presents the observations and the observational limits on IRS~13E's infrared variability. Sec.~3 describes models quantifying the expected variability in different scenarios.  Sec.~4 addresses GC sources other than IRS~13E, and Sec.~5 summarizes this paper's results.
\begin{figure*}
  \centering
\includegraphics[width=0.49\linewidth]{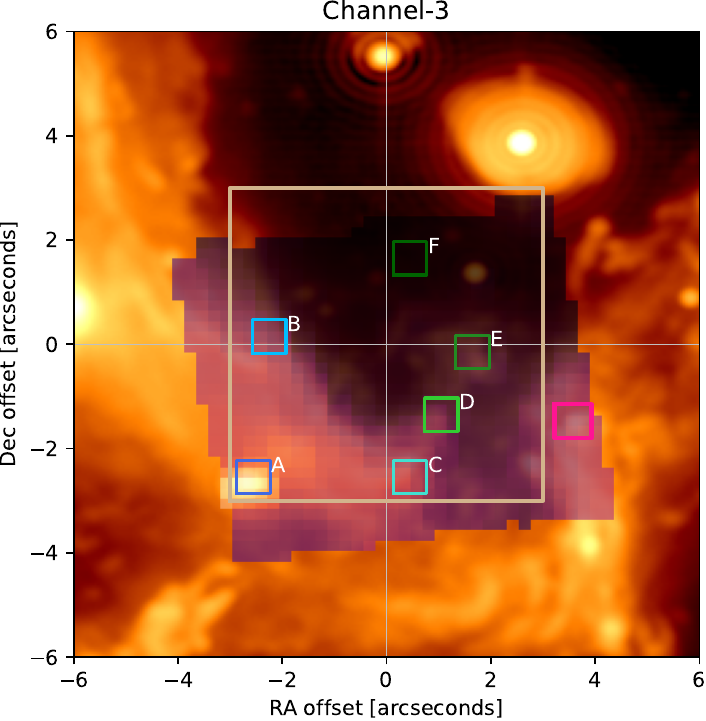}
\includegraphics[width=0.49\linewidth]{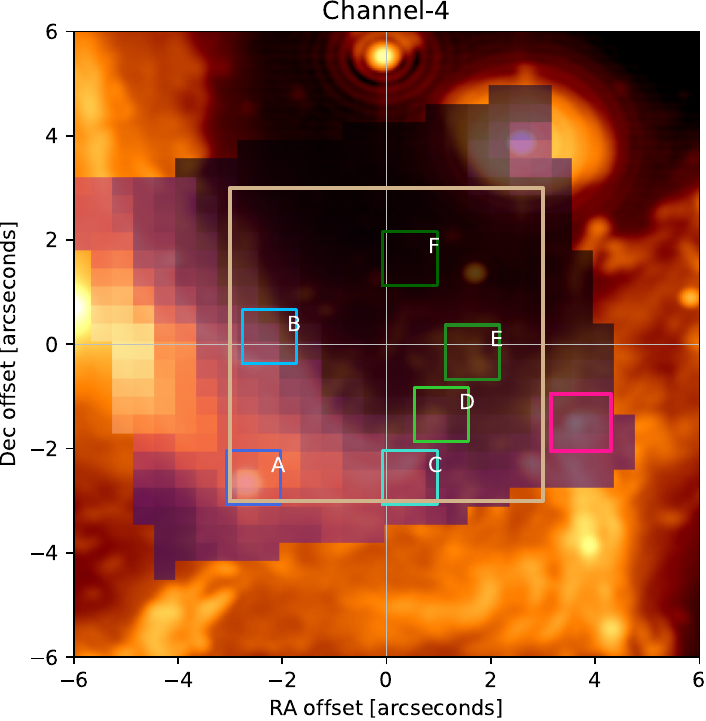}
\caption{JWST/MIRI images of channels 3 (${\sim} 12.3~\mu$m, left) and 4 (${\sim} 19.5~\mu$m, right) overlaid on the VISIR  8.6 $\mu$m image  \citep{Dinh2024}. IRS~13E is visible as a diffuse and moderately bright object marked by pink boxes in the lower right corner of each MIRI image. Regions are marked A--F are reference regions ordered in decreasing brightness.  The box sizes indicate the $3\times 3$ pixel apertures used for extracting fluxes. The channel~4 pixels are nearly a factor of two larger than the the channel~3 pixels, and the aperture centres are not exactly aligned, but the pink boxes still enclose the brightest region of IRS~13E\null. The large orange square is $6''\times6''$ in size and indicates the approximate MIRI FoV\null. 
}
\label{fig:channel-fov}
\end{figure*}

\section{Constraints on IMBH variability}
\label{sec:data_processing}
\subsection{Observations \& Data Calibration}
JWST observations were obtained as part of Program 4572 (Cycle 2 GO, PIs: D. Haggard, J. Hora, G. Witzel). The observations used the MIRI Medium Resolution Spectrograph \citep[MRS;][]{MIRI2015, MIRI2023}. MIRI/MRS allows simultaneous observations in four non-contiguous mid-infrared (MIR) wavelength bands (``channels'') with the channel wavelengths depending on which of three grating angles was used.\footnote{\url{https://jwst-docs.stsci.edu/jwst-mid-infrared-instrument/miri-observing-modes/miri-medium-resolution-spectroscopy}} The time-series ``staring'' observations reported here used the SHORT grating angle and were centred on \Sg. IRS~13E's offset from \Sg\ is 3\farcs2 in right ascension, $-1\farcs51$ in declination \citep{Dinh2024}, and IRS~13E was inside the field of view only in the two longest-wavelength channels, 11.55–-13.47 and 17.70–-20.95~\micron. The data were calibrated with the standard JWST pipeline version 1.17 \citep{jwst_pipeline_bushouse2024} in context \texttt{pmap}  1322. The calibration used the same procedure as  \cite{Fellenberg_2025}, and  their paper gives details of generating the time-resolved spectral cubes.
In the reduced data, the pixel sizes were 0\farcs2 and 0\farcs35 in the two channels. Owing to the crowded environment in the GC and the associated bright thermal emission, the modest angular resolution of the MIRI IFU does not resolve any individual members of the IRS~13E cluster. However, the cluster is visible as a diffuse object in both channels as shown in \autoref{fig:channel-fov}. 

\subsection{IRS~13E Light Curve Generation}\label{ssec:lc_gen}

The data set comprises 11 staring observations obtained on 2024 April 4, 6, 8, and 9 and September 6. Each observation lasted ${\sim}$2 hours giving ${\sim} 22$~hours on the source. Each observation gave a time-resolved spectral cube at 86-second cadence over the $\sim$2 hours. Strong spectral lines, primarily  [Ne II] 12.8~$\mu$m and [S III] 18.8~$\mu$m had to be masked. We binned the spectral axis in channels 3 and 4 into three equal intervals ($\Delta\lambda = 0.65~\mu$m for channel 3 and $\Delta\lambda=1.08~\mu$m) to increase the signal-to-noise during each integration. Individual bin boundaries and central frequencies/wavelengths are given in Table~\ref{tab:channel-bins}. 

\tamojeet{Additionally, the MRS IFU data usually suffers from a non-negligible degree of fringing \citep{Law_2023, Law_2024} in both spatial and wavelength/spectral axes, and could potentially vary temporally if the pointing drifts. This is partly taken care of by turning on the optional residual de-fringing step in the reduction pipeline. For spatial and spectral fringing, we sum the flux in a $3\times 3$ aperture in RA/Dec, and average the flux in wavelength bins of width 0.6--1 $\mu$m. This averages out the fringing and makes the relative variations go down considerably. To a minor extent, it affects the photometric calibration, but our uncertainty in the final flux estimate is dominated by the uncertainty in the extinction correction ($\delta m\approx 0.2 ~\mathrm{mag}$). Temporally, minor shifts in pointing could cause changes in fringing, but we verify that the pointing is stable to $<250 \mu$as for our observations \citep{Michail_SED}. Further, the dominant effect of the pointing drift is the change in the actual flux level itself, which we account for by subtracting a linear drift.} 

\begin{table*}[ht]
\caption{Wavelength Bins and Correction Factors for IRS~13E Light Curves}
\centering
\begin{tabular}{ccccc}
\hline\hline
Central & Central & Wavelength & Extinction & Aperture \\
Frequency (THz) & Wavelength ($\mu$m) & Boundaries ($\mu$m) & Correction ($m$) & Correction ($k$) \\
\hline
25.3 & 11.87 & 11.55--12.19 & 1.26 & 2.01\\
24.0 & 12.51 & 12.19--12.83 & 0.68 & 1.99\\
22.8 & 13.15 & 12.83--13.47 & 0.52 & 2.06\\
16.4 & 18.24 & 17.69--18.77 & 1.36 & 1.56\\
15.5 & 19.32 & 18.77--19.86 & 1.17 & 1.53\\
14.7 & 20.40 & 19.86--20.94 & 0.86 & 1.47\\ \hline
\end{tabular}
\raggedright\null\\
{\sc Note}: the extinction corrections $m$ \citep[from][]{Fellenberg_ext}  are in magnitudes, and the\\ aperture corrections $k$ \citep[from Appendix~B of][]{Michail_SED} are multiplicative factors. 
\label{tab:channel-bins}
\end{table*}

The region around IRS~13E is dominated by diffuse emission from both IRS~13E itself and from nearby dust. For the goal of measuring source variability, we summed the flux in a $3\times3$-pixel window (0\farcs6 in channel~3 and 1\farcs05 in channel~4, each approximately $1.5\times$ the FWHM at their respective wavelengths; \citealt{Law_2023}) centred on the source. These beam sizes are the largest permitted by the field of view (Fig.~\ref{fig:channel-fov}), but they still provide reliable estimates of the flux around the putative IMBH location. The crowded field at JWST's angular resolution means other sources are included in the beam, and the beam is too small to capture all of IRS~13E's flux. Therefore the extracted fluxes are not reliable as to the absolute flux of IRS~13E alone. 

To set meaningful limits on variability, the observed light curves have to be corrected for flux outside the $3\times3$-pixel aperture (``aperture correction'' $k$) and for interstellar extinction ($m$). The aperture correction has some uncertainty because of the non-dithered nature of the observations, and
Appendix~B of \cite{Michail_SED} describes how the correction was derived.  In short, the star 10 Lacertae was measured in a 72-point dither pattern using the chosen aperture, and the $3\times3$-pixel measurements were scaled to the ``infinite-aperture'' flux in a single image made from all the dithers. 

The extinction corrections were taken directly from Fig.~5 of \citet{Fellenberg_ext}. Table~\ref{tab:channel-bins} gives both correction factors for each wavelength bin.
After applying these corrections, given in Table~\ref{tab:channel-bins}, to each of the six wavelength bins, we averaged the channel~3 and~4 results to produce light curves for each observation in each channel. To look for short-timescale variability, we then subtracted the median of each observation and detrended it with a best-fit linear function. Finally, we measured the standard deviation of all of the detrended light curves.

\begin{figure}[h!]
  \centering
  \includegraphics[width=\columnwidth]{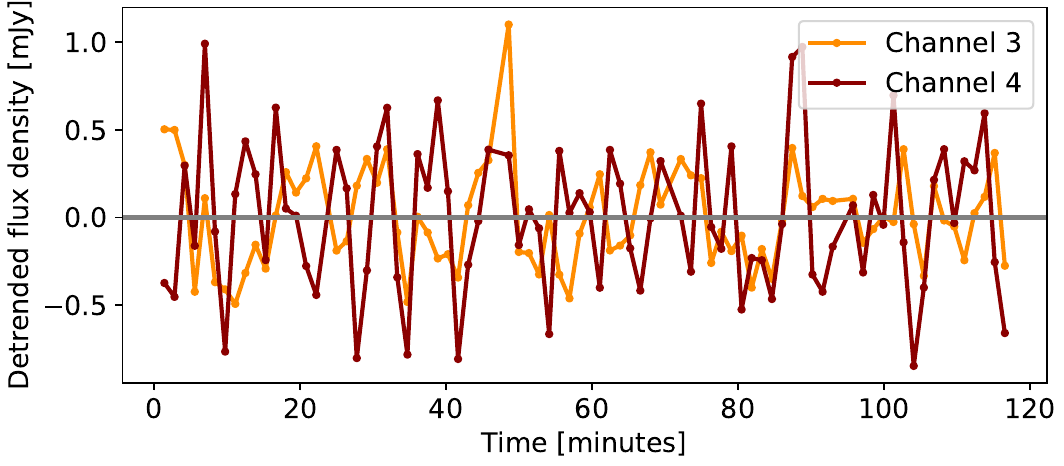}
  \caption{Sample light curve of IRS~13E\null. Coloured points show residual flux densities for channels~3 and~4 as indicated in the legend.  The residuals were calculated by subtracting the median flux density of~2859 and 5394~mJy for the two channels, respectively and removing a linear drift. The light curve is from April~6 exposure~2, the same epoch where \Sg\ showed a mid-infrared flare \citep{Fellenberg_2025}. Unlike the \Sg\ light curve, there is no correlation between the two channels.}
  \label{fig:sample-lc}
\end{figure}

Fig.~\ref{fig:sample-lc} shows one light curve. The two channel lightcurves do not show any flare-like features at or above the $3\sigma$ level. Further, they show no correlation between channels and are dominated by photometric noise, consistent with the absence of intrinsic variability. We also inspect inspect residual maps as a function of time and there was no detection of a point source in those (unlike the Sgr A* flare in \citealt{Fellenberg_2025}, Figure 2) in any of the exposures.

The uncertainty $e$ on any given upper-limit flux $u$ must take into account the photometric uncertainty, the uncertainty of the aperture-correction factor, and the uncertainty in the extinction correction. Let the photometric time series be denoted $\{l_t\}$ with $t$ running from $1$ to $T$. Assume this has a standard deviation of $\sigma_{\rm{obs}}$. Also let each measurement have uncertainty $\{\delta l_t\}$. Then, using the extinction correction $m$ (with uncertainty $\delta m$) and aperture correction $k$ (with uncertainty $\delta k$) for that wavelength, we have the intrinsic variability and the upper limit flux $u$ given by
\begin{align}
\sigma_{\rm{true}} &= \sqrt{\sigma_{\rm{obs}^2}- \langle \delta l_t^2 \rangle}\\
u &= 3\sigma_{\rm{true}}\cdot10^{0.4m} \cdot k\quad,
\end{align}
and the uncertainty $e$ is defined as
\begin{align}
    \delta\sigma &= \sqrt{\sum_{t=1}^T\left(\frac{l_t}{T\sigma} \right)^2 \left(\delta l_t\right)^2} 
    \end{align}
    \begin{align}
    e &= 3\cdot10^{0.4m}\sqrt{(k\cdot\delta\sigma)^2 + (0.921k\sigma\cdot\delta m)^2 + (\sigma\cdot\delta k)^2}
\end{align}
with suitable multiplicative constant factors for unit conversions as needed. 

These give us $u \pm e$ as an upper limit on the intrinsic flux variability at that position. We note that even in the absence of any point source being resolved at the position of IRS 13E, the upper limit on the flux variation of the full aperture flux also acts as the upper limit on the flux variation of IRS 13E alone -- since any other variable objects in the aperture would only increase the variation of the total aperture flux.

Further, in the following section, we use this flux variability upper limit to compute an absolute flux upper limit. Like the variability, the absolute flux upper limit for the full aperture also acts as an upper limit for the putative IMBH's flux alone, despite no clear point source being resolved. Once an observed absolute flux upper limit is established, we can rule out various models if their flux predictions are in excess of the observed upper limit flux.

\begin{figure*}[h!]
  \centering
  \includegraphics[width=0.75\linewidth]{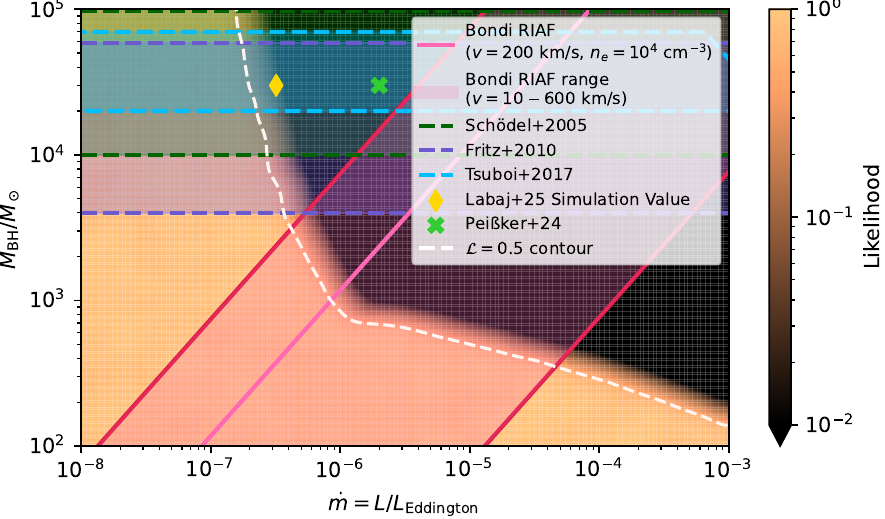}
  \\
  \includegraphics[width=0.75\linewidth]{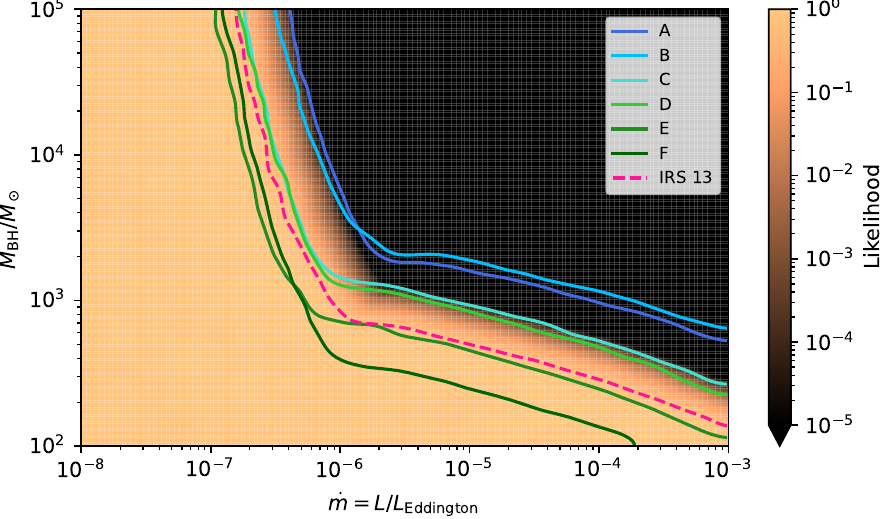}
  \caption{Likelihoods of $M_{\rm{BH}}$,$\dot{m}$ combinations imposed by the variability upper limits. The upper panel is for IRS~13E, and the lower one is for all  regions in the field of view. Both are smoothed using a Gaussian kernel of width $0.02 \times$ the span of each parameter. Likelihoods  are shown according to the respective colour bars, and the dashed line in each panel (white top panel, red bottom panel) shows the $\mathcal{L}=0.5$ contour for IRS~13E\null.  In the upper panel, the green cross marks the  proposed black-hole mass and accretion rate of \cite{Peissker_2024}, and  the yellow diamond shows the accretion rate derived by \cite{Labaj_2025arXiv} from MHD simulation of stellar winds from six WR stars in the IRS~13E cluster. The stellar mass is the value used by both authors. Allowed mass ranges from previous studies are marked by pairs of dashed lines: green, purple, and blue for \cite{Schoedel2005, Fritz2010, Tsuboi2017}, respectively. The nominal Bondi accretion regime is indicated by a solid pink line, and the range of reasonable parameters is between the solid red lines. In the lower panel, solid lines show contours of $\mathcal{L}=0.5$  for the six representative regions (A-–F), and background shading shows the combined likelihood. 
  }
  \label{fig:double-likelihood}
\end{figure*}

\section{Results and Discussion}

\begin{table*}[ht]
\centering
\caption{Median Flux Densities and Upper Limits on Variability} 
\begin{tabular}{cccccccccccc}
\hline\hline\rule{0pt}{4.5ex}
Region & \multicolumn{1}{c}{\shortstack{Position \\ RA, Dec}} & \multicolumn{2}{c}{\shortstack{Channel 3 \\ flux density (mJy)}} 
       & \multicolumn{2}{c}{\shortstack{Channel 3 \\ variation (mJy)}} 
       & \shortstack{Channel 3\\luminosity}
       & \multicolumn{2}{c}{\shortstack{Channel 4 \\ flux density (mJy)}} 
       & \multicolumn{2}{c}{\shortstack{Channel 4 \\ variation (mJy)}} 
       & \shortstack{Channel 4\\luminosity}\\ 
       & \tamojeet{[arcsec]} & Obs. & Corr. & Obs. & Corr. &  $10^{32}$ ergs s$^{-1}$ & Obs. & Corr. & Obs. & Corr. & $10^{32}$ ergs s$^{-1}$\\
\hline\rule{0pt}{3.ex}
IRS 13E & $-3.2,-1.5$ & 2859 & 10643 & 0.34 & 1.27 &24& 5394 & 24243 &  0.45 & 2.02 & 25\\
A       & $2.6,-2.6$& 5646 & 21018 & 6.29 & 23.42 &440& 15174 & 68201  & 4.16 & 18.70\0 & 230 \\
B       & $2.3,0.1$& 3439 & 12802 & 3.18 & 11.84 &220& 9131 & 41040  & 1.68 & 7.55 & 93 \\
C     & $-0.4,-2.6$  & 2797 & 10412 & 0.64 & 2.38 &45& 8773 & 39431  & 0.75 & 3.37 & 41 \\
D     & $-1,-1.4$  & 1633 & 6079  & 0.26 & 0.97 &18& 4087 & 18369  & 0.47 & 2.11 & 26\\
E     & $-1.6,-0.2$  & 1096 & 4080  & 0.32 & 1.19 &22& 2223 & 9991   & 0.38 & 1.71 & 21\\
F     & $-0.4,1.6$  & 176  & 655   & 0.11 & 0.41 &7.8& 661  & 2970   & 0.23 & 1.03 & 13\\
\hline
\end{tabular}
\raggedright\null\\[1ex]
{\sc Note}: variations are 3$\sigma$ upper limits. The rows show IRS 13E and the six regions shown in Fig.~\ref{fig:channel-fov}. Positions are offsets from \Sg\ in arcseconds. Values are for a $3\times 3$-pixel aperture before (``Obs.'') and after (``Corr.'') applying  extinction and aperture corrections. Luminosities are $\nu L_\nu$ of the corrected upper limit at the Galactic Centre's distance. \tamojeet{Sample lightcurves of each region from the 6th April 2024 dataset can be found at \url{https://github.com/tamojeetroychowdhury/IRS-13E-JWST}. Lightcurves for any of the exposures may be generated using GO 4572 data from MAST and following the steps outlined in Sec.~\ref{sec:data_processing}.}
\label{tab:flux_vars}
\end{table*}

\subsection{Expected IMBH variability}
While the details of accretion-flow physics are a field of active research \citep[e.g.,][]{Dexter2009,Moscibrodzka2013,Ripperda2022}, the basic spectral energy distribution (SED) is well captured by simple semi-analytic models \citep[e.g.,][]{Yuan2003}, but can also be modelled with emission from a jet-base \citep[e.g.,][]{Falcke1995, Falcke2004}. The latter describe sources better, if they feature prominant radio jets \citep{Ontiveros2023}. For Sgr A*-like accretion, the effective differences between the models are small, which is why we focus on the models of the first kind; based on the observational constraints, our calculations could be trivially repeated for any other appropriate model.

One prediction of such one-dimensional RIAFs is that the gyro frequency of the synchrotron emission, $\nu_b$ in terms of $M_{\rm{BH}}$ and the Eddington ratio $\dot m \equiv \dot M / \dot M_{\rm{Edd}}$, is given by \citep{Pesce_2021}: 
\begin{multline}
    \nu_b\approx 
    (1+\beta)^{-1/2}\alpha^{-1/2}c^{-1/2}_1 c^{1/2}_3 M_{\rm{BH}}^{-1/2}\dot{m}^{1/2}r^{-5/4+s/2}
    \\ \times[4\times10^{15}~\mathrm{Hz}]\quad,\label{eq:riaf_peak}
\end{multline}
where $c_1\approx0.5$ and $c_3\approx0.3$ are constants defined by \cite{Narayan1995}, $\alpha$ is the accretion disk viscosity, $r$ is the radius, $\beta \approx 10$ is the ratio of gas pressure to magnetic pressure, and $s\approx-0.8$ is the radial power law index of the accretion flow's density. Further, it is shown that the SED peak frequency depends on the mass and accretion rate of the MBH\null. 
For typical $M_{\rm{IMBH}}=10^{3}$--$10^{5}~\mathrm{M_{\odot}}$) and RIAF accretion rates ($\dot{m}_{\rm{IMBH}}=10^{-9}$--$10^{-4}$ assuming that the plasma properties are similar to that around \Sg), the SED peaks in the MIR\null.

\Sg, with $M_{\rm BH}=4\times10^6$~\Msol\ \citep[e.g.,][]{Boehle2016}, has $\nu_b$ at sub-mm wavelengths. \Sg's variability at those wavelengths is well characterized and occurs on timescales of tens of minutes to hours \citep[e.g.,][]{Dexter2014,eht_sgra_I}.\footnote{This is the timescale of \Sg's sub-mm variability, not the flaring events seen in the X-ray and infrared. Flares likely originate from non-thermal emission processes and have different temporal characteristics at different emission frequencies \citep[e.g.,][]{vonFellenberg2023_sgra,vonFellenberg2024}} The variability timescale of any RIAF accretion flow (assuming sub-Eddington accretion) is expected to scale linearly with mass \citep[e.g.,][]{Gammie2003}, and therefore the variability timescale $t_{\rm{var}}$ of a $10^3~\mathrm{M_\odot}$ IMBH is in the range of less than a minute to a few minutes. Further, Eq~\ref{eq:riaf_peak} shows that while the SED peak of \Sg\ occurs in sub-mm wavelengths, the SED peak for an IMBH would occur in the MIR.

\Sg's variability amplitude at $\nu_b$ is typically  $\sigma_{\rm{\Sg}}\approx 20\%$--50\%, \citep[e.g.,][]{Dexter2014, VonFellenberg2018,Wielgus2022}. Therefore a conservative limit for a RIAF IMBH is $\sigma_{\rm{IMBH}}>10\%$. This choice is also theoretically sound, as similar variability values are recovered in (mass-scale-free) general relativistic magnetohydrodynamic (GRMHD) simulations of accretion flows targeting \Sg\ and its surrounding environment \citep[e.g.,][]{eht_paper_V}. If the variability is $>$10\% of intrinsic flux, that flux must be no larger than 10$\times$ the  variability upper limit. This constraint is, at most wavelengths, 2--3 orders of magnitude tighter than obtained from directly measuring the SED owing to crowding, with the measured flux coming from unresolved  multiple sources.

\subsection{Previous Constraints on IRS 13E}
For a putative IMBH to bind the IRS~13E cluster stars, its mass would have to be ${\ga} 10^4$~\Msol\ \citep{Schoedel2005}. This mass could also explain ALMA observations of ionised gas in an eccentric orbit \citep{Tsuboi2017} and the high-eccentricity orbits of a subset of the ``clockwise stars'' in the Galaxy's central 0.5~pc \citep{Zheng2020}. At  $10^4$~\Msol, the Eddington ratio is ${\la} 2\times 10^{-7}$ corresponding to a  mass-accretion rate ${\la}4.5 \times 10^{-10}~\rm M_\odot $~yr$^{-1}$. Further, at the distance of IRS~13E from \Sg, a proper motion analysis by \cite{Reid2020} ruled out any IMBH more massive than ${\sim} 10^{4.8}$~\Msol. While this mass is close to the upper limit of the parameter space we are exploring already, the proper-motion constraint means there is no need to consider higher masses.

Recently, \cite{Peissker_2024} inferred the presence of an IMBH at IRS 13E with a mass $\sim 3\times 10^4\ M_\odot$ from kinematic analysis of stellar orbits in the central few arcseconds and Eddington ratio $\dot m \approx 2\times10^{-1}$ using a broad-band SED. Also, \cite{Labaj_2025arXiv} used simulations of six Wolf--Rayet (WR) stars in an IRS 13E--like physical configuration and wind-fed accretion to derive an Eddington ratio of $\dot m \approx 3.17 \times 10^{-7}$ for a typical IMBH mass of  $\sim 3\times 10^4\ M_\odot$. We show in subsequent subsections and in Fig~\ref{fig:double-likelihood} that both these scenarios can be ruled out.

\subsection{IMBH RIAF Models}
The \cite{Pesce_2021} RIAF model predicts the expected range of flux densities  in the observed frequency bins. The model's main parameters are the black hole mass $M_{\rm{BH}}$ and the Eddington ratio $\dot m$. For the present computations, all other parameters of the model, such as the density scaling $n_e\propto r^{s}$, $\alpha$, $c_1$, and $c_3$ were left at the default values. We used the Python package \texttt{dynesty} \citep{Speagle2020_dynesty} to sample the parameter space using log-uniform priors on $M_{\rm{IMBH}} \in \left[10^2, 10^5\right]~\mathrm{M_\odot}$ and $\dot{m} \in \left[10^{-8}, 10^{-3}\right] $.
The model provides a predicted flux density $f(\nu)$ for each $M_{\rm{BH}}$, $\dot m$ combination.

The likelihood $\mathcal{L}$ for the combination of six observed upper limits (one for each wavelength bin) on flux density $u_i$ and uncertainties $e_i$ is given by
\begin{align}
    \mathcal{L}(p) = \prod_{\rm{all \ epochs}}\prod_{i=1}^6 \Phi\left( \frac{u_i-f_i}{e_i} \right)\quad,
\end{align}
where $f_i$ are the calculated values and $\Phi$ is the cumulative distribution function of a  zero-mean, unit-variance Gaussian distribution.
The likelihood was computed over the full parameter space using $50\,000$ samples and is shown in Figure~\ref{fig:double-likelihood}.  Figure~\ref{fig:double-likelihood} also shows two previous suggestions for possible black-hole accretion rates.

\subsection{Bondi Accretion}
For a black hole accreting from the ambient interstellar medium (ISM), as would be the case for IRS~13E, the Bondi mass-accretion rate is given by \citep[][their Eq.~2]{Yutaka_1998}:
\begin{equation}
\begin{split}
    \dot M \approx & ~7.4 \times 10^{13}\ \mathrm{g\ s}^{-1}~\times \\ & \left( \frac{M_{\rm{BH}}}{\Msol} \right)^2 \cdot \left( \frac{n}{10^2\ \rm{cm}^{-3}} \right) \cdot \left( \frac{v}{10\ \rm{km\ s}^{-1}} \right)^{-3},
\end{split}
\end{equation}
where $n$ is the typical particle density in the environment, and $v$ is the typical velocity dispersion of gas.  In the Mini-Spiral and the Western Arm, $n\approx10^4~\mathrm{cm^{-3}}$ \citep{Ferriere_2012, Zhao2009}. A typical velocity dispersion in the GC is $v\approx200~\mathrm{km~ s^{-1}}$ \citep{Genzel2003}, but values in the range 10--600~km~s$^{-1}$ are possible, where the lower value is given by the sound speed of a $\sim10000~\mathrm{K}$ plasma. The \Sg\ RIAF is observed to be accreting at  $10^{-5}$ of the Bondi rate \citep{Quataert2002}, which can also be derived theoretically \citep[][]{Baganoff2003_spectrum,Genzel2010} assuming a universal scale of the accretion density $n_e(r) \propto r^{-1}$ and $\dot{m}(r)\propto^{-0.5 \dots -1}$. 
While it is unclear whether this scaling holds for any accreting system in the GC, specifically if the accreting gas has non-zero angular momentum at the Bondi radius, consistent scaling relations have been found in RIAF accretion simulations for \Sg\ \citep{Ressler_2018, Ressler2020} fed by Wolf--Rayet (WR) stars from large distances and explicitly for an IRS~13E--like system \citep{Labaj_2025arXiv}. Similar scaling relations has also been argued to hold universally for low Eddington rate accreting black-hole systems \citep[e.g.,][]{Narayan1994,Abramowicz1995,Blandford1999,Pang2011,Xu2023}. 

Recently, \cite{Lalakos2025} provided a scaling ration of the Bondi accretion rate to a magnetically-arrested disk (MAD) RIAF which was derived from large scale MHD simulations ($\sim 10^4~\mathrm{R_g}$), for which they claim universality:
\begin{align}
    \frac{\dot M_{\mathrm{BH}}}{\dot M_\mathrm{Bondi}} = 10^{-3} \times \left( \frac{r_\mathrm{B}}{10^5 r_\mathrm{g}} \right)^{-0.66}
\end{align}
which gives a factor of $\sim 1.3 \times 10^{-3}$ for $v_\mathrm{eff}=200$ km s$^{-1}$. 
The corresponding expression in terms of the Eddington ratio gives $\dot m \propto M_{\rm{BH}}$. The allowed range is indicated in \autoref{fig:double-likelihood}, which shows a substantial overlap between the expected RIAF accretion range and the parameters allowed by IRS~13E's SED, albeit primarily in the low-mass regime which has been ruled out by previous dynamical constraints.

\subsection{Constraints from Future Radio Observations}
We compute the rough observable radio (and X-ray) fluxes that would be observed for the range of values of $M_{\mathrm{IMBH}}$ and $\dot m$. We follow similar steps as \cite{Gaggero2017}, and compute $L_\mathrm{bol} = \eta \dot M c^2$ where $\eta = 0.1\dot M / \dot M_{\mathrm{Edd}}$. Further we use $L_X = 0.3L_{\mathrm{bol}}$ following \cite{Fender2013}, giving us the X-ray luminosity. We do not convert this to an expected counts/s as it involves assumptions about the spectrum and instrument characterisation.

To go from the X-ray luminosity $L_X$ to the radio luminosity $L_R$, we use the fundamental plane relation from \cite{Plotkin2012}:
\begin{align}
    \log L_X = & (1.45 \pm 0.04) \log L_R - (0.88 \pm 0.06) \log M_{\rm{BH}} - (6.07 \pm 1.10)
\end{align}
For a $10^3 M_\odot$ IMBH accreting at $10^{-6}$ times the Eddington rate, this gives a radio luminosity $\nu L_\nu\sim 5\times 10^{25}$ erg s$^{-1}$. This corresponds to about 100 nJy at 5 GHz, at which point most radio facilities are confusion-limited for the GC regions. Future telescopes such as the Square Kilometre Array (SKA) may be able to reach these limits with very deep integrations \citep{Schodel_SKA}.

Higher masses and accretion rates can provide higher values of radio flux that would be more readily detectable, but these values can already be ruled out at the sensitivity levels of our current JWST MIRI observations.

\section{Extending to the full field of view}
\subsection{Selected regions around the \Ms}
The above procedure for measuring variability can be used in any region that has time-series photometry, and the resulting upper limit constrains the parameters of any putative black hole within the field of view. 
\DEL{We create a grid in the range of $[-3'', 3'']$ offset from the location of \Sg\ in steps of $0.3''$ covering the central $6''\times6''$ of the GC. The regions are chosen so that each is covered at least once in channels 3 or 4. We repeat the process to create variance-based SEDs using the standard deviation of the time-series in the same fashion as before and deriving the likelihood estimated over the parameter space of $(M_{\rm{BH}}, \dot m)$.}

\autoref{fig:double-likelihood} shows the results for six representative $3\times 3$-pixel regions marked in \autoref{fig:channel-fov}. Point~A, the brightest, overlaps with IRS~21, and point~F, the faintest, is outside the \Ms\ in a low-background region. Other points on or around the \Ms\ were selected based on visual assessment of relative brightness. 

The  $\mathcal{L}=0.5$ contours have similar shapes at all surface brightnesses, but as expected the limits on BH mass and accretion rate are tighter in  areas with lower surface brightness.
Large black holes could be hidden if they are not accreting, but that is not a reasonable scenario in the high-density GC region. The photometric upper limits are given for the representative regions in Table~\ref{tab:flux_vars}. The observed flux variability scales with the observed absolute flux of the region (i.e., highest for A, lowest for F)\null. Because there are no point source detections, and we have large apertures in a crowded field, the observed variability is treated as an upper limit rather than the true value of the intrinsic variability in a putative IMBH\null. In general, even for the largest variations (in the brightest regions), the observations rule out any IMBH more massive than ${\approx} 2\times10^{3}$~\Msol\ if it is accreting at ${\geq} 10^{-6}$ times the Eddington rate  (Fig.~\ref{fig:double-likelihood}).

\subsection{Full Field of View around \Sg}
\begin{figure}
  \centering
\includegraphics[width=\columnwidth]{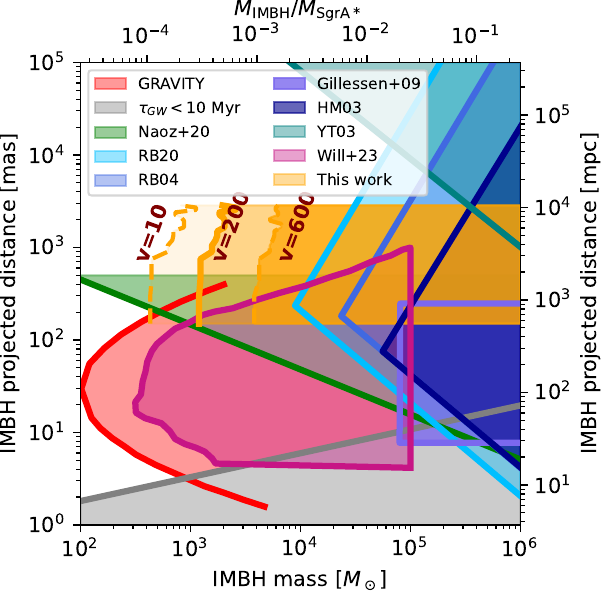}
\caption{Galactic Centre IMBH parameter ranges excluded by observations. IMBH mass is shown horizontally and  projected distance from \Sg\ vertically, and shaded areas indicate excluded ranges. This figure is based on Figure~D2 of \citet{GRAVITYCollaboration_schwarzschild}, and our new exclusions based on Bondi RIAF accretion are shown in orange. The differing zones for IMBH velocities of 10, 200, and 600~km~s$^{-1}$ are marked. Other constraints are from \citet[][labeled Naoz+20 in the legend]{Naoz2020}, \citet[][Will+23]{Will2023}, \citet[][RB04 and RB20]{Reid2004, Reid2020}, \citet[][Gillessen+09]{Gillessen2009}, \citet[][HM03]{Hansen2003}, and \citet[][YT03]{Yu2003}.  The grey area at bottom is excluded by the gravitational-wave inspiral timescale \citep{GRAVITYCollaboration_schwarzschild}.}
\label{fig:mr}
\end{figure}

To constrain the maximum mass of an IMBH anywhere in the field of view, we examined $3\times 3$-pixel regions in the channel~3 and~4 images at radii between 0\farcs15 and 3\arcsec\ from \Sg. At each radius (in steps of 0\farcs15), we used the pixel with the highest variability to compute the maximum IMBH mass at that radius. Because variability increases with flux density, the pixel with the maximum variability is in most cases located in the \Ms.
To obtain the mass constraint, we assumed the Bondi-to-RIAF accretion rate, which for a given effective velocity and particle density provides an accretion rate as function of mass $\dot{m}_{\rm{IMBH}}(M_{\rm{IMBH}})$. This gave constrained slices in ($M_{\rm{IMBH}},\dot{m}_{\rm{IMBH}}$) space. To derive the constraints, we assumed a typical velocity dispersion in the GC of $200~\mathrm{km~s^{-1}}$ \citep[e.g.,][]{Genzel2003_stars} and a particle density in the \Ms\ of $10^4$ cm$^{-3}$ \citep[e.g.,][]{Zhao2009}. 
We also included the two extremal cases: a maximally co-moving IMBH, where the minimal velocity is given by the approximate sound speed for a $10^4~\mathrm{K}$ plasma ($v_{\rm{min}}=10$ km s$^{-1}$) and an IMBH moving in the opposite sense of the \Ms\ IMBH ($v_{\rm{max}}=600$ km s$^{-1}$). Fig.~\ref{fig:mr} shows the allowed range of IMBH masses as a function of radial distance. 

All photometric constraints are highly sensitive to the effective velocity of the IMBH\null. This is a trivial consequence of the $\dot{m}\propto v^{-3}$ dependence of the accretion rate. While contra-moving IMBHs can evade photometric detection, the observations place tight constraints on co-moving IMBHs. The most stringent constraints (i.e., in regions~E and~F, having the least intrinsic brightness and the least variabiltiy) reach sub-$10^3~\mathrm{M_\odot}$ values. These sensitivities put the most massive merger products ($\sim$190--265~\Msol; \citealt{Ligo260Msun2025}) discovered by LIGO--VIRGO--KARGA within reach. Such ``IMBHs'' peak at shorter wavelengths and have shorter variability timescales, perhaps making JWST/NIRCam more suitable to constrain them. Recent observations with  18~s integrations at $\lambda_{\rm{obs}}=4.5~\mathrm{\mu m}$ \citep[][]{Yusef-Zadeh2025} reach ${\sim} 10^{-5}~\mathrm{Jy}
$. The better angular resolution at near-infrared wavelengths compared to MIR would also lead to tighter constraints. 

\section{Conclusions}
The absence of significant photometric variability of IRS~13E constrains the combination of ($M_{\rm{IMBH}}$, $\dot{m}_{\rm{IMBH}}$)\null. The constraints are tight enough to rule out two potential IMBH candidates: one observationally proposed \citep{Peissker_2024} and one simulation-based scenario of a wind-fed IMBH \citep{Labaj_2025arXiv}. In the latter case, the specific choice of a $3\times 10^4~\mathrm{M_\odot}$ IMBH firmly is ruled out, but it is conceivable that slightly lighter IMBHs (that also satisfy the lower limits on mass from dynamical constraints like those of \citealt{Schoedel2005}) may be present in IRS~13E.

The full ${\sim} 6''\times 6''$ field of view around \Sg\ also shows no evidence for an IMBH\null. A scaling relation for Bondi-to-RIAF accretion rules out a large parameter space for higher accretion-rate IMBHs. Such accretion rates occur if the IMBH moves at speeds similar to the GC velocity dispersion, and in this case  $M_{\rm{IMBH}} < 10^4~\mathrm{M_\odot}$. Counter-moving IMBHs are expected to have lower accretion rates and can therefore evade detection more easily. For them, the constraint is looser, $M_{\rm{IMBH}} < 10^5~\mathrm{M_\odot}$. In the other direction, a putative IMBH that co-moves with the gas would have a high accretion rate, and photometric sensitives can limit such BHs to masses less than a few hundred~$\mathrm{M_\odot}$.

The constraints reported here rely on simple Bondi--Hoyle accretion theory, which we have scaled to mimic RIAF accretion. The scaling is motivated observationally by accretion onto \Sg\ \citep[e.g.,][]{Genzel2010,Ciurlo2025} and theoretically by accretion simulations onto \Sg\ \citep{Ressler_2018}, a putative IMBH in an IRS~13E--like system \citep{Labaj_2025arXiv}, and general accretion simulations \citep[e.g.,][]{Pang2011,Xu2023}. Despite these justifications, the importance of the question warrants dedicated simulations of accretion from ambient gas onto compact objects in the GC (Labaj et al., in prep.).

Our observations demonstrate that photometric IMBH constraints are feasible, and these JWST/MIRI observations constrain an important parameter space for the first time. 
Tantalizingly, masses not far below the co-moving detection limit have recently been observed by LIGO--VIRGO--KARGA with a derived occurrence rate of order $10^{-4}~\mathrm{M_\odot ~Gpc~yr^{-1}}$ \citep[][]{LigoMassDistribution2023,Ligo260Msun2025}.

Galactic Centre observations with current instruments are confusion limited and therefore cannot reach their nominal photometric sensitivities. This will change for the next generation of extremely large telescopes (ELTs). Specifically, the first light imager MICADO at the ESO ELT should reach photometric sensitivity $F_{{\rm limit}, K}\approx29~\mathrm{mag} \approx 4~\mathrm{nJy}$ \citep[][]{Davies_2021} corresponding to de-extincted luminosity at the Galactic Centre of 
${\approx} 10^{30} ~\mathrm{erg~s^{-1}}$. For the Galactic Centre, MICADO should be confusion-limited only in the central arcseconds, where the stellar density of ${\sim} 10^4/\rm{arcsecond^2}$ will limit the photometric sensitivity to $K\approx25$  \citep{Bordoni2025, Sturm2024arXiv240816396S}. Beyond the central area, the stellar density drops steeply. 
Thus direct photometric detection of stellar-mass black holes residing in the Galactic Centre will be within MICADO's reach.

\begin{acknowledgements}
We thank Rainer Sch{\"o}del for providing the VISIR image used in \autoref{fig:channel-fov}, which was originally published by \cite{Dinh2024}.
We thank Matus Labaj, Sean Ressler, Michal Zajacek, and Bart Ripperda for the helpful discussions on accretion in the GC\null.
We thank Aditya Vijaykumar for the useful feedback on the gravitational wave event GW231123 and the gravitational-wave merger products.
This research was supported by the International Space Science Institute (ISSI) in Bern, through ISSI International Team project \#24-610, and we thank the ISSI team for their generous hospitality.

SDvF gratefully acknowledges the support of the Alexander von Humboldt Foundation through a Feodor Lynen Fellowship and thanks CITA for their hospitality and collaboration.

SDvF, BR, BSG are supported by the Natural Sciences \& Engineering Research Council of Canada (NSERC), the Canadian Space Agency (23JWGO2A01), and by a grant from the Simons Foundation (MP-SCMPS-00001470). BR acknowledges a guest researcher position at the Flatiron Institute, supported by the Simons Foundation.

JM is supported by an NSF Astronomy and Astrophysics Postdoctoral Fellowship under award AST-2401752. 

SM is supported by a European Research Council Synergy Grant ``Blackholistic'' (grant 10107164).

DH, ZS, NMF acknowledge support from the Canadian Space Agency (23JWGO2A01), the Natural Sciences and Engineering Research Council of Canada (NSERC) Discovery Grant program, the Canada Research Chairs (CRC) program, the Fondes de Recherche Nature et Technologies (FRQNT) Centre de recherche en astrophysique du Québec, and the Trottier Space Institute at McGill.
NMF acknowledges funding from the FRQNT Doctoral Research Scholarship.

TR acknowledges funding support from the Deutscher Akademischer Austauschdienst (DAAD) Working Internships 
\end{acknowledgements}

\bibliography{irs13e}{}
\bibliographystyle{aasjournalv7}

\end{document}